\documentclass[twocolumn,aps,pra]{revtex4}
\usepackage[latin9]{inputenc}
\usepackage{amsmath}
\usepackage{amssymb}
\usepackage{graphicx}

\makeatletter
\@ifundefined{textcolor}{}
{%
 \definecolor{BLACK}{gray}{0}
 \definecolor{WHITE}{gray}{1}
 \definecolor{RED}{rgb}{1,0,0}
 \definecolor{GREEN}{rgb}{0,1,0}
 \definecolor{BLUE}{rgb}{0,0,1}
 \definecolor{CYAN}{cmyk}{1,0,0,0}
 \definecolor{MAGENTA}{cmyk}{0,1,0,0}
 \definecolor{YELLOW}{cmyk}{0,0,1,0}
}

\usepackage{upgreek}

\@ifundefined{textcolor}{}{%
\definecolor{BLACK}{gray}{0}
\definecolor{WHITE}{gray}{1}
\definecolor{RED}{rgb}{1,0,0}
\definecolor{GREEN}{rgb}{0,1,0}
\definecolor{BLUE}{rgb}{0,0,1}
\definecolor{CYAN}{cmyk}{1,0,0,0}
\definecolor{MAGENTA}{cmyk}{0,1,0,0}
\definecolor{YELLOW}{cmyk}{0,0,1,0}
}

\usepackage{color}%\bibliographystyle{plain}
\def\NOT(#1,#2){\OneQubitGate(#1,#2){$X$}}

\@ifundefined{showcaptionsetup}{}{%
 \PassOptionsToPackage{caption=false}{subfig}}
\usepackage{subfig}
\makeatother

\begin{document}

\title{Polarizing the electronic and nuclear spin of the NV-center in diamond
in arbitrary magnetic fields: analysis of the optical pumping process}

\author{Tanmoy Chakraborty, Jingfu Zhang and Dieter Suter}

\affiliation{Fakult�t Physik, Technische Universit�t Dortmund, D-44221 Dortmund,
Germany}
\begin{abstract}
Initializing a set of qubits to a given quantum state is a basic prerequisite
for the physical implementation of quantum-information protocols.
Here, we discuss the polarization of the electronic and nuclear spin
in a single Nitrogen vacancy center in diamond. Our initialization
scheme uses a sequence of laser, microwave and radio-frequency pulses,
and we optimize the pumping parameters of the laser pulse. A rate
equation model is formulated that explains the effect of the laser
pulse on the spin system. We have experimentally determined the population
of the relevant spin states as a function of the duration of the laser
pulse by measuring Rabi oscillations and Ramsey-type free-induction
decays. The experimental data have been analyzed to determine the
pumping rates of the rate equation model. 
\end{abstract}

\maketitle

\section{Introduction}

The nitrogen-vacancy (NV) center in diamond \cite{Aharonovich} has
been identified as an excellent solid state quantum bit system, which
provides the possibility of implementing quantum protocols at room
temperature \cite{Jelezko_PRL,Dutta_Science,Neumann_Science}. Most
of these applications use the long coherence times of the NV center
\cite{Balasubramanian_NatMater} and the optical initialization and
readout of the system \cite{Chu_2015,Yale_2013,Jelezko_review}. The
NV center has been used for several interesting demonstrations like
robust multipartite entanglement persisting over a timescale of milliseconds
at room temperature \cite{Neumann_Science}, quantum interference
between photons \cite{Bernien_PRL,Sipahigil_2012}, an efficient quantum
memory \cite{Barrett_2005,Stoneham_2009,Shim_2013}, single shot readout
of single nuclear spins \cite{Robledo_Nature} or quantum gate operations
where dephasing is protected with the help of dynamical decoupling
\cite{Jingfu_2015}. These developments of quantum information processing
based on the defect centers in diamond have remarkably boosted solid
state quantum technology and pioneered a new way towards reliable
implementation of quantum computation \cite{Maurer_2012}.

For diamond crystals with low density of spins \cite{Gaebel_Nature},
it is possible to find single NV centers which remain magnetically
well isolated from other defect centers. An NV center, consisting
of the electronic spin and one or several nuclear spins, is therefore
an interesting physical realization of a quantum register. Local control
of the targeted qubits within a single NV center can be performed
by specifically addressing the concerned center and manipulating the
system with laser pulses, microwave (MW) and radio-frequency (RF)
magnetic fields \cite{Childress_Science}. \cite{PhysRevLett.116.043603,APL(105)242402}.
Nuclear spins are useful resources for storing and transmitting quantum
information \cite{kane1998silicon,gershenfeld1997bulk}. It is necessary
to enhance the strength of nuclear spin polarization since it improves
the efficiency of detection of nuclear magnetic resonance (NMR) signal
and hence increases the signal-to-noise ratio. In this context, DNP
(dynamic nuclear polarization)\cite{overhauser1953polarization,abragam1978principles}
has been widely established as an effective technique which relies
on transferring the spin polarization from electrons to nuclei. Optically
induced DNP \cite{sukhenko1985calculations,tateishi2014room} has
been successfully applied to polarize the nuclear spins in systems
like diamond \cite{king2015room,scheuer2016optically,alvarez2015local}
and silicon carbide \cite{falk2015optical}. Other novel approaches
include high fidelity nuclear spin initialization and single shot
read out at excited state level anticrossing\cite{dreau2013single,pla2013high,NeumannScience542},
implementation of Hartmann-Hahn double resonance technique\cite{PhysRevLett.111.067601,scheuer2016optically},
population trapping protocols \cite{APL(105)242402,Shim_2013,PhysRevLett.116.043603}
etc. However, the resulting polarization using certain pulse sequences
\cite{APL(105)242402,Shim_2013,PhysRevLett.116.043603}, like a scheme
presented through this paper, are limited by the fact that the initialization
technique in the beginning using a long laser pulse although can create
a strong electronic polarization, it cannot completely initialize
the system into the $m_{S}=0$ state \cite{NewJ.Phys.13.025013(2011)}.
On the other hand, such techniques have the advantage that these experiments
do not require any specific magnetic field value.

In the present work, we discus the initialization of a quantum register,
which is an essential prerequisite for the implementation of a universal
quantum computer \cite{DiVincenzo_2000}. In particular, we investigate
a protocol for polarizing the electronic and the $^{14}$N nuclear
spin of a single NV center in diamond. We have employed a sequence
of pulses which is quite similar to what has been described in Ref.
\cite{APL(105)242402}, where Pagliero et al. have employed MW, RF
and laser pulses to polarize the nuclear spins associated with NV
centers. However, the primary focus of our paper is optimizing the
pulse sequence, evaluating the optical control parameters for the
spins which play a substantial role in this initialization scheme
and determining the purity of the relevant individual states by performing
partial state tomography. For this purpose, we introduce a rate equation
model which can explain the influence of optical irradiation on the
spins and provide a physical picture of population transfer between
different quantum states of the system. Thus, determination of the
laser pumping parameters by analyzing the experimental data will help
in initializing a quantum register to a target state in a more deterministic
and optimized way. While the initialization of the electronic spin
is a standard element of all applications of the NV center, it is
more difficult to control the nuclear spin. To tackle this issue,
the excited state level anti-crossing in a magnetic field has been
used \cite{Jacques_2009,PhysRevB.81.035205}. The present technique
does not rely on the anti-crossing and can therefore be applied at
all strengths and orientations of the magnetic field. The sequence
starts with the usual initialization of the electronic spin by a laser
pulse. We then swap the electronic and the nuclear spin and apply
a second laser pulse to re-initialize the electron spin. Since the
optical initialization procedure partly depolarizes the nuclear spin,
it is important to understand the dynamics of the coupled spin system
and to optimize the sequence such that the purity of the targeted
state is maximized.

In the following, we discuss the system of interest and describe the
initialization scheme. We then analyze the dynamics of the system
in the presence of non-resonant optical irradiation and use the results
for an optimal state preparation.

\section{System and Setup}

A single NV center embedded in a $^{12}$C-enriched (concentration
of 99.995\%) diamond sample was chosen as the experimental system.
The effective dephasing time $T_{2}^{*}$ for this center is $\sim40\mu s$,
as measured by a Ramsey-type free-induction decay experiment. The
Hamiltonian of the system consisting of an electronic spin ($S$=1)
coupled to a $^{14}$N nuclear spin ($I$=1) is

\begin{equation}
\mathcal{H}=DS_{z}^{2}-\gamma_{e}BS_{z}+PI_{z}^{2}-\gamma_{n}BI_{z}+AS_{z}I_{z},\label{eq:Hamiltonian}
\end{equation}
where $D$=2.87 GHz is the zero-field splitting, $S_{z}$ and $I_{z}$
are the $z$-components of the electronic and nuclear spin operators,
$A$ =-2.16 MHz is the hyperfine coupling, $\gamma_{n}$=3.1 MHz/T
and $\gamma_{e}$ =-28 GHz/T are the nuclear and electronic gyromagnetic
ratios and $P$ =-4.95 MHz represents the nuclear quadrupole coupling.
Here, we assume that the magnetic field is oriented along the NV symmetry
axis.

The measurements were performed at room temperature on a single NV
center embedded in a diamond crystal. A diode-pumped solid-state laser
emitting green light at 532 nm was employed for exciting the NV center
in a home-built confocal microscope. The CW laser beam was passed
through an acousto-optic modulator (AOM) with a rise time of 50 ns
and an extinction ratio of 58 dB to generate pulses. The beam was
then focused on a single NV center with a microscope objective (numerical
aperture =1.4) mounted on a nano-positioning system. The optical power
at the sample was $\sim150\mu$W. A MW signal generator (APSIN) and
a direct digital synthesis (DDS) radio-frequency (RF)-source were
used to generate the microwave (MW) signal, which was subsequently
passed through a switch and an amplifier to create the MW pulses with
suitable frequency, amplitude and phase for manipulating the electronic
spins. Another DDS, switch and amplifier were used to generate the
RF pulses for manipulating the nuclear spins. The MW and the RF pulses
were passed through a combiner and a Cu wire attached to the surface
of the sample. A permanent magnet generated a magnetic field of 6.1
mT along the axis of the NV center.

\section{Initialization Procedure and The RateEquation Model}

\begin{figure}[h]
\centering{}\includegraphics[width=1\columnwidth]{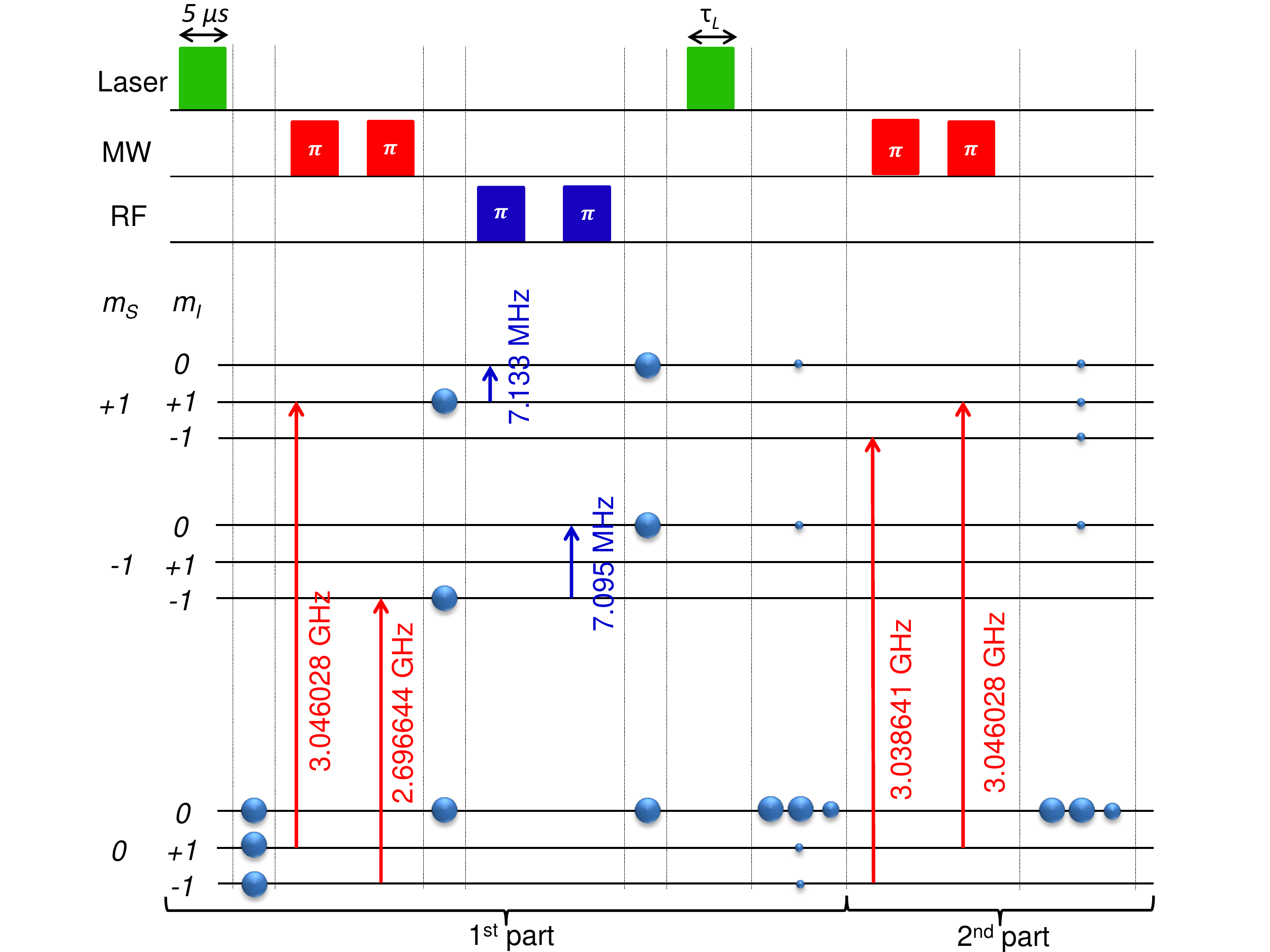}\caption{Pulse sequence for initializing the system into the $\left|0,0\right\rangle $
state. The occupation probability of the energy states have been schematically
represented by the size of the filled circles. MW = microwave, RF
= radio frequency. The time interval between the laser, MW and RF
pulses were $\approx2\mu s$ which hardly affects the initialization
scheme as the $T_{1}$ for both the electronic and nuclear spin are
$\sim ms$. \label{fig:pulses}}
\end{figure}

Fig. \ref{fig:pulses} shows the eigenstates and the transition frequencies
between the relevant energy levels. To illustrate the pulse sequence
in a convenient way, we list the states $\left|m_{S},m_{I}\right\rangle $
in the following order ($m_{S},m_{I}$) = (0,-1; 0,+1; 0,0; -1,-1;
-1,+1; -1, 0; +1,-1; +1,+1; +1,0) from the bottom to the top in the
energy level scheme of Fig. \ref{fig:pulses} (the actual sequence
of energies is slightly different), and in subsequent parts of the
article.

The initialization process, which is also represented in Fig. \ref{fig:pulses}
starts with a long (5 $\mathrm{\mu}$s) laser pulse, which initializes
the electronic spin into the bright ($m_{S}=0$) state, while the
nuclear spin is fully depolarized. As mentioned earlier, it cannot
completely initialize the system into the $m_{S}=0$ state and the
actual polarization is usually unknown where different values are
found in the literature. However, for the purpose of this paper, it
is sufficient to determine the populations relative to the initial
population of the $m_{S}=0$ state. Accordingly, conservation of populations
implies that the sum of all populations must always be unity, $\sum_{i=1}^{9}P_{i}=1$
. Also, we consider only the electronic ground state of the NV$^{-}$
system, since the excited state population as well as the population
of the NV$^{0}$ states are small under our experimental conditions
(excitation with 532 nm laser light) \cite{Aslam_NJP,Waldherr_PhysRevLett.106.157601}.
After the 5 $\mathrm{\mu}$s laser pulse, two MW $\pi$ pulses swap
the populations of the $\left|0,+1\right\rangle $ $\leftrightarrow$
$\left|+1,+1\right\rangle $ and the $\left|0,-1\right\rangle $ $\leftrightarrow$
$\left|-1,-1\right\rangle $ states. Their pulse durations were 3.874
$\mu$s and 1.456 $\mu$s, respectively. Two RF pulses swap the populations
between the nuclear spin states $\left|+1,+1\right\rangle $ $\leftrightarrow$
$\left|+1,0\right\rangle $ and $\left|-1,-1\right\rangle $ $\leftrightarrow$
$\left|-1,0\right\rangle $; both pulses had the same duration of
105.908 $\mu$s. This sequence of four pulses thus exchanges the polarizations
of the electronic and the nuclear spin: the nuclear spin becomes fully
polarized in the $m_{I}=0$ state, but the electron spin becomes fully
depolarized. The electronic spin therefore has to be re-polarized
by a second laser pulse. Since this laser pulse affects not only the
electronic spin, but also the nuclear spin, we have to analyze its
effect on the spin system and optimize its duration. The $2^{nd}$
part of Fig. \ref{fig:pulses} shows an additional purification step
that can be used to remove population from the ground state that is
not in the $m_{I}=0$ nuclear spin state.

We consider the effect of the laser on the spin system as a simple
redistribution of populations between the different eigenstates of
the spin Hamiltonian. We write $\vec{P}$ for the populations of the
nine states and use the sequence of states defined in Fig. \ref{fig:pulses}.
Writing $k_{S}$ for the rate with which the electron spin changes
from the $m_{S}=\pm1$ state to the $m_{S}=0$ state and $k_{I}$
for the rate at which the nuclear spin flips between any pair of states,
the equation of motion can be expressed in matrix form as 
\begin{equation}
\frac{d}{dt}\vec{P}=M(k_{S},k_{I})\vec{P},\label{eq:RateEqn}
\end{equation}
with the full form of the matrix $M$ given in the appendix.

Starting from the state $\vec{P}=\frac{1}{3}(1,1,1,0,0,0,0,0,0)$
generated by the first laser pulse, the MW and RF pulses generate
the population vector $\vec{P}=\frac{1}{3}(0,0,1,0,0,1,0,0,1)$. Starting
from this initial condition, the formal solution of Eq. (\ref{eq:RateEqn})
for the evolution of the populations during the second laser pulse
is

\begin{eqnarray}
\vec{P} & = & \frac{1}{3}(1-e^{-k_{S}\uptau_{L}}\frac{2k_{I}}{(3k_{I}-k_{S})}-e^{-3k_{I}\uptau_{L}}\frac{(k_{I}-k_{S})}{(3k_{I}-k_{S})},\nonumber \\
 &  & 1-e^{-k_{S}\uptau_{L}}\frac{2k_{I}}{(3k_{I}-k_{S})}-e^{-3k_{I}\uptau_{L}}\frac{(k_{I}-k_{S})}{(3k_{I}-k_{S})},\nonumber \\
 &  & 1-e^{-k_{S}\uptau_{L}}\frac{2(k_{I}-k_{S})}{(3k_{I}-k_{S})}+e^{-3k_{I}\uptau_{L}}\frac{2(k_{I}-k_{S})}{(3k_{I}-k_{S})},\nonumber \\
 &  & 0,0,e^{-k_{S}\uptau_{L}},0,0,e^{-k_{S}\uptau_{L}}).\label{eq:Population_Eqn}
\end{eqnarray}
The general solution for arbitrary initial conditions is given in
the appendix.

\begin{figure}[h]
\centering\includegraphics[scale=0.3]{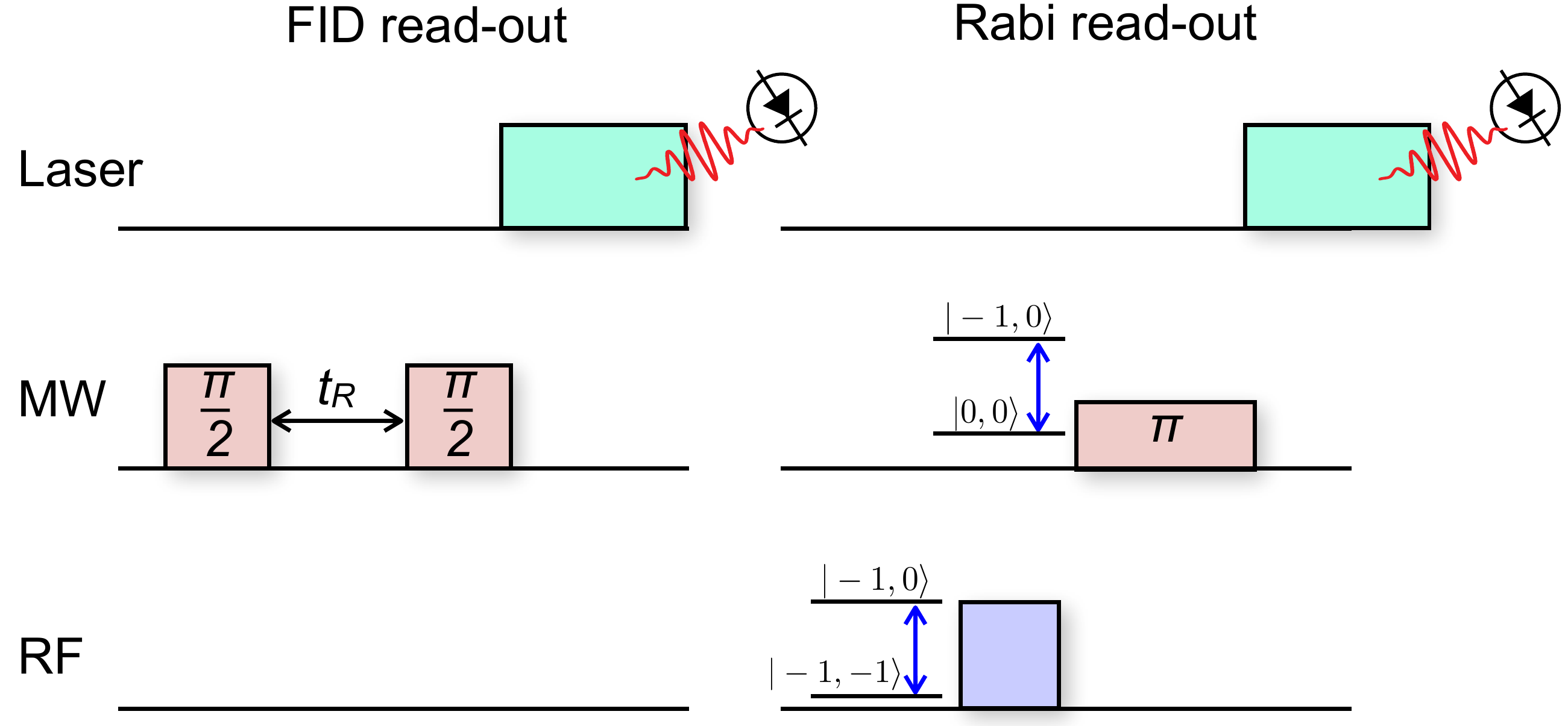}\caption{Pulse sequences for measuring the free-induction decay of the electronic
spin and the nuclear spin Rabi oscillation. For the FID scheme, the
hard MW pulses are non-selective for all transitions from the $m_{S}=0$
to the $m_{S}=-1$ state of the electron spin. In the Rabi scheme,
the pulses are selective for the single transitions indicated in the
figure. \label{fig:pulse_readout}}
\end{figure}

\section{Experimental results and discussion}

To test and optimize the preparation scheme outlined above, we performed
experiments for different durations of the second laser pulse and
analyzed the resulting state by performing partial quantum state tomography.
For this purpose, we measured Ramsey-type free induction decays, using
the sequence shown in figure \ref{fig:pulse_readout}. The pulse sequence
for this measurement consists of two identical $\frac{\pi}{2}$ MW
pulses separated by the free evolution time $t_{R}$. The first MW
pulse generates a coherent superposition of the states $m_{S}=0$
and $m_{S}=-1$ which subsequently evolves for a time $t_{R}$. The
second MW pulse converts the coherence into population difference
which is eventually measured by the read-out laser pulse. The detuning
frequency $\nu_{det}$, determines the phase $\phi$ of the second
$\frac{\pi}{2}$ MW pulse by $\phi=\phi_{c}+2\pi\nu_{det}t_{R}$ where
the constant $\phi_{c}$ is offset from the actual electronic transition
frequency. A Fourier transform of the time-domain data generates the
frequency-domain spectra. During these measurements we applied hard
$\frac{\pi}{2}$ MW pulses to excite all the three electron spin transitions
corresponding to nuclear spin states $m_{I}=0,\pm1$ between electronic
spin states $m_{S}=0$ and $-1$.

\begin{figure}[h]
\centering{}\includegraphics[scale=0.3]{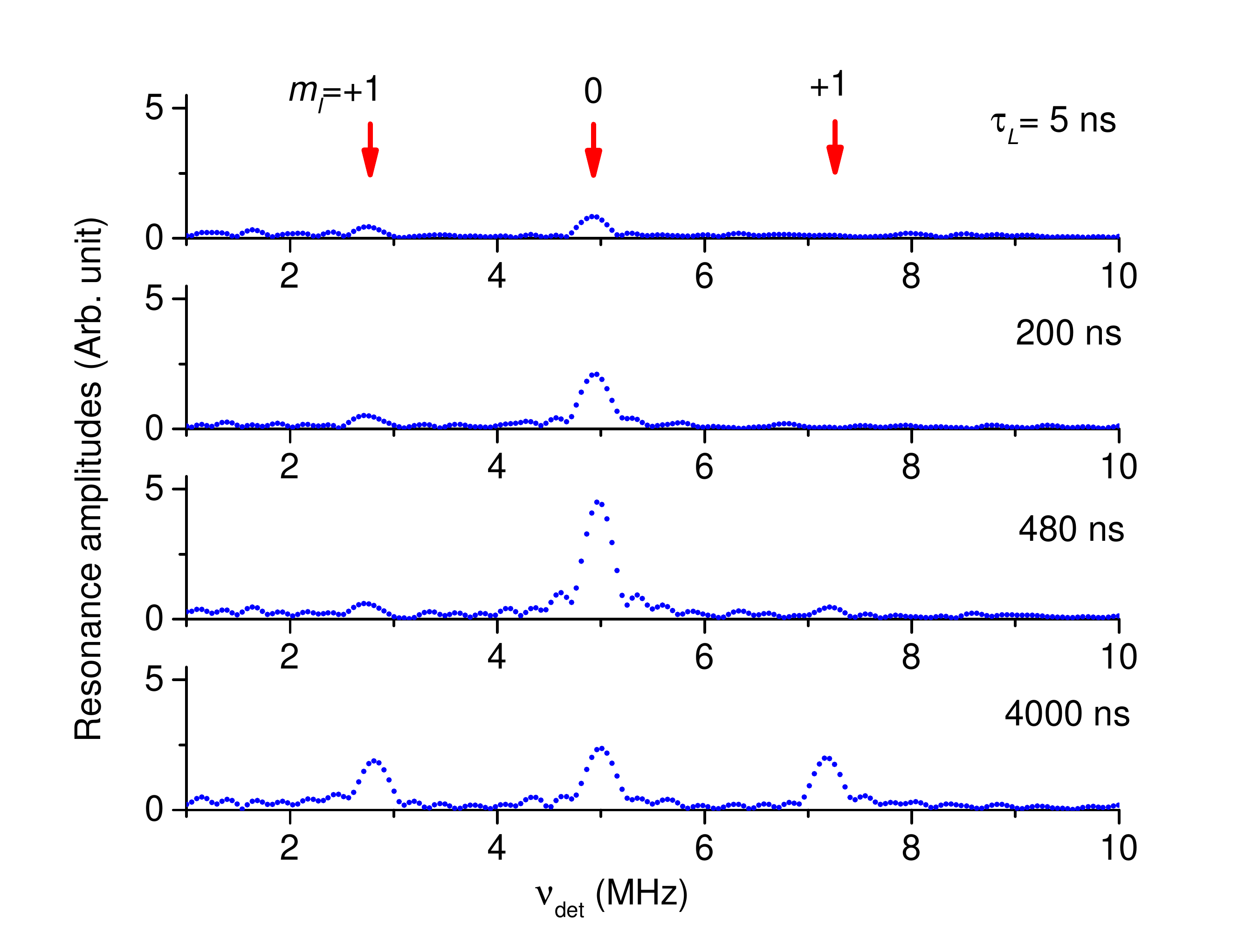}\caption{ESR spectra, measured as Fourier transforms of Ramsey fringes for
four different values of the laser pulse duration $\uptau_{L}$. The
amplitudes are proportional to the population difference between the
$m_{S}=0$ and $m_{S}=-1$ state for the three different nuclear spin
states. The initial condition before applying the laser pulse with
duration $\uptau_{L}$ was $\vec{P}=\frac{1}{3}(0,0,1,0,0,1,0,0,1)$.
The Ramsey measurements were done for free evolution times of up to
4 $\mu$s, which limits the widths of the resonance lines in the spectra.
\label{fig:spectra}}
\end{figure}

Figure \ref{fig:spectra} shows four of the resulting spectra measured
after the second laser pulse for pulse durations $\uptau_{L}$ = 5ns,
200ns, 480ns and 4000ns. The arrows in the figure indicate the nuclear
spin states to which the resonance lines are associated. Their amplitudes
$A_{k}$ are proportional to the population differences between the
$m_{S}=0$ ground state of the electron spin and the $m_{S}=-1$ state:
$A_{k}=P_{0,k}-P_{-1,k}$, where $k=0,\pm1$ indicates the nuclear
spin quantum number. 

\begin{figure}[h]
\centering\includegraphics[scale=0.3]{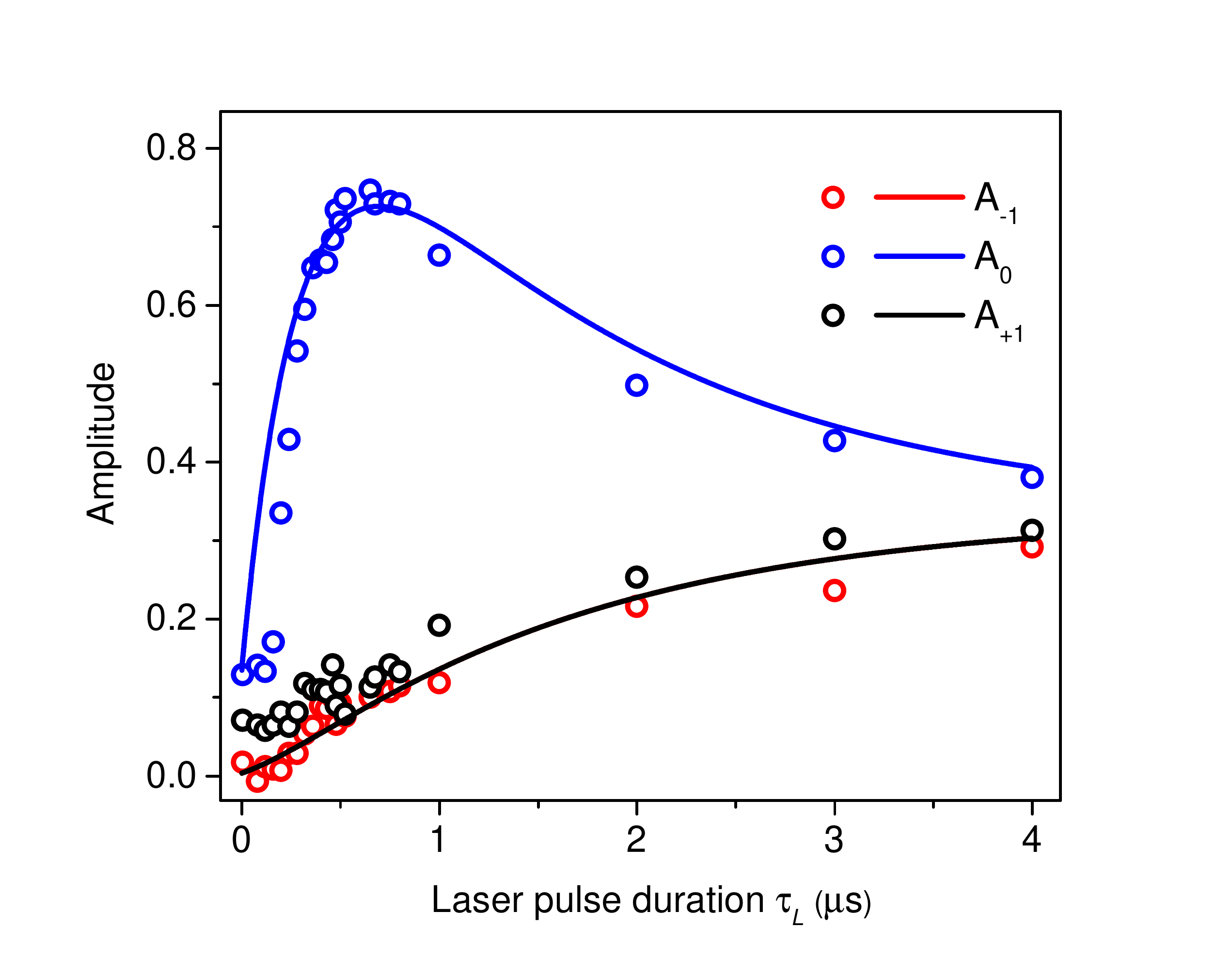}\caption{Measured amplitudes of the resonance lines for the nuclear spin states
$\left|-1\right\rangle $, $\left|+1\right\rangle $, and $\left|0\right\rangle $
and their sum $A+A_{+1}+A_{0}$ (as mentioned in the graph legend)
along with the corresponding theoretical curves as described in Eq.
\ref{eq:Population_Eqn}. \label{fig:amplitudes}}
\end{figure}

Fig. \ref{fig:amplitudes} shows the measured amplitudes as a function
of the laser pulse duration. For short laser pulses, all population
differences vanish. For long durations, the populations of the three
ground states $|0,0\rangle$ and $|0,\pm1\rangle$ converge to 1/3,
while the $m_{S}=-1$ population vanishes, resulting in population
differences of 1/3. The experimental data are well consistent with
the theoretical curves plotted using Eq. \eqref{eq:Population_Eqn}.

\begin{figure}[h]
\subfloat[]{\includegraphics[width=42mm]{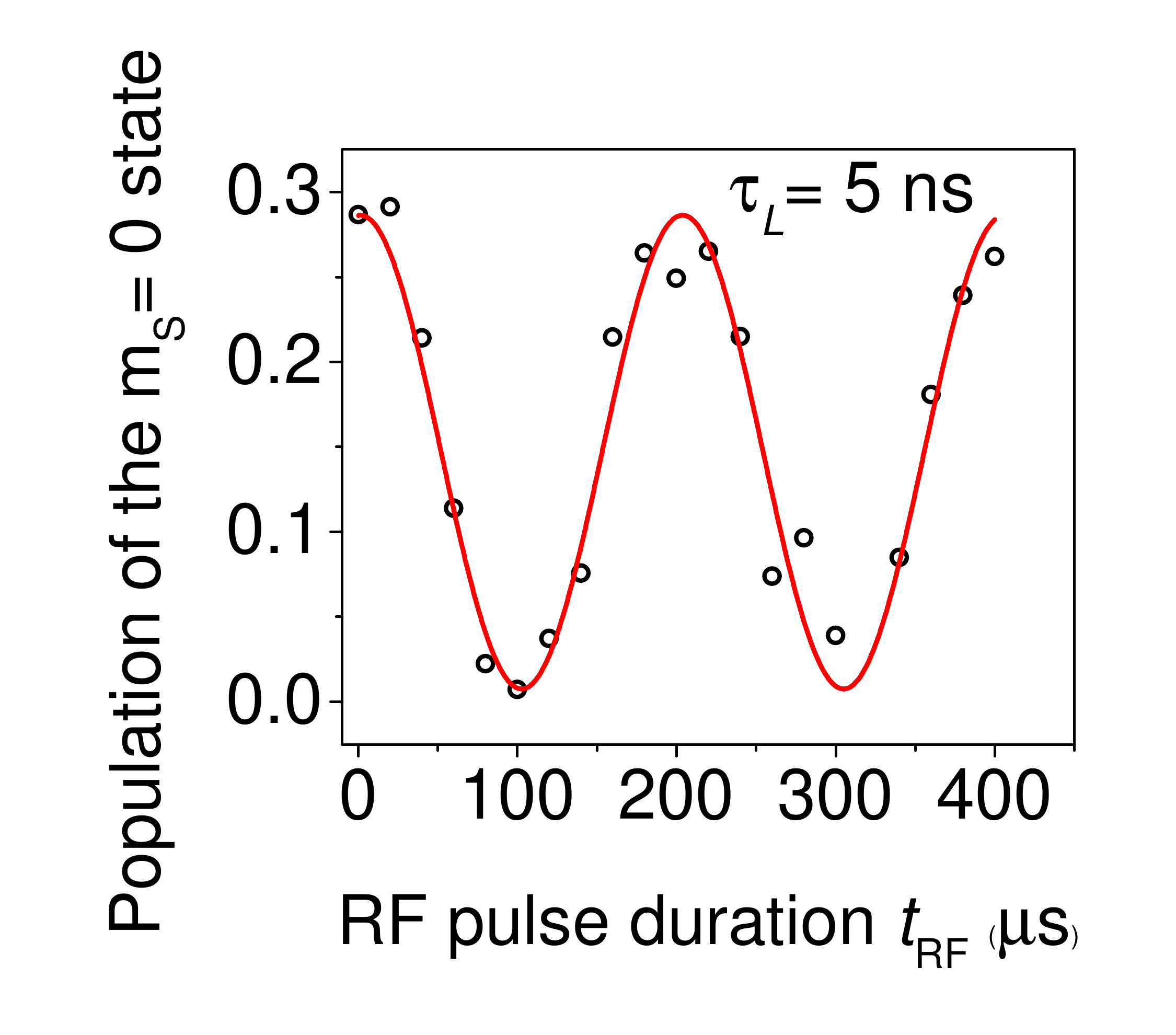}}\subfloat[]{\includegraphics[width=42mm]{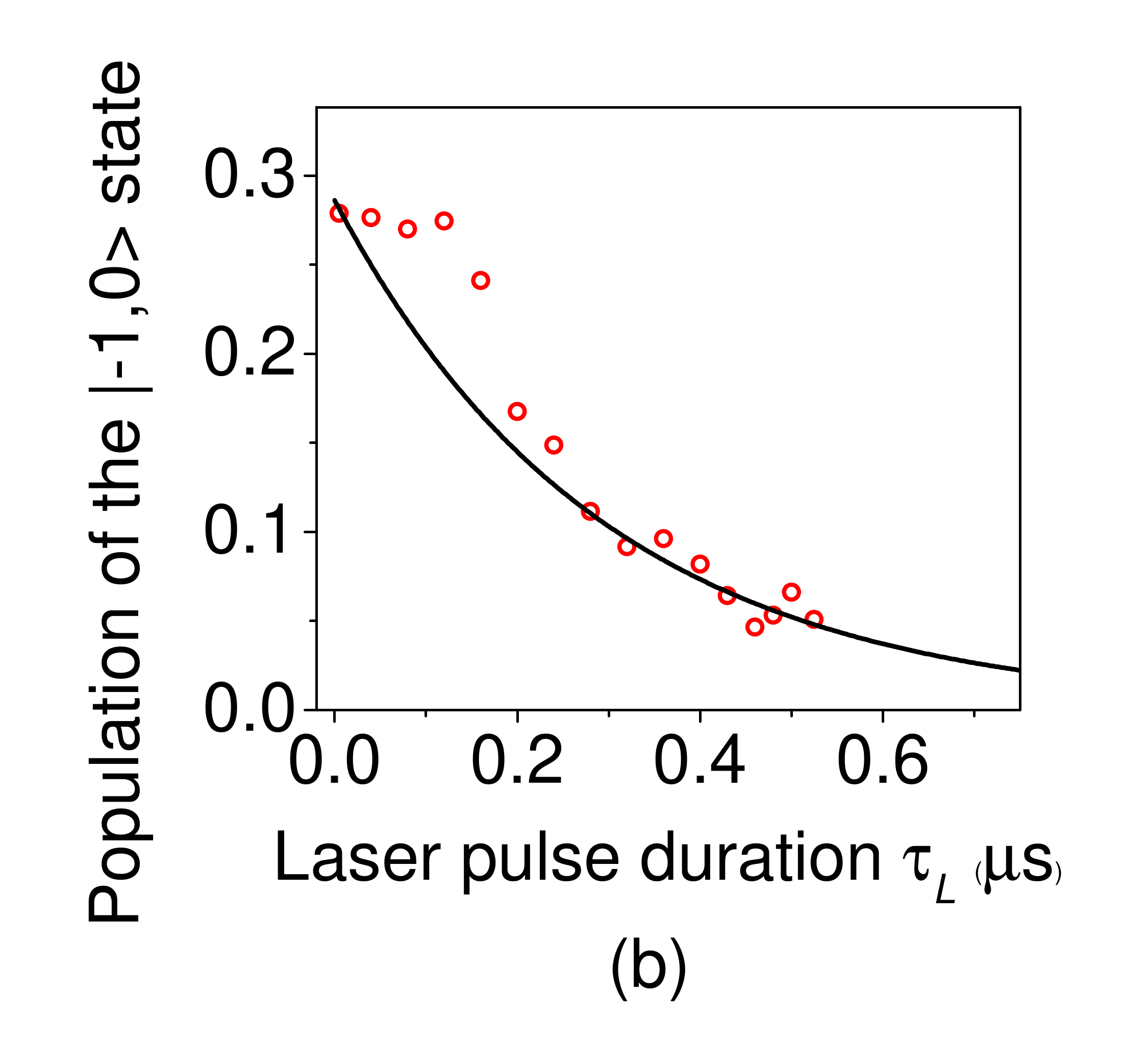}}\caption{(a) Rabi oscillation data between the $\left|-1,-1\right\rangle $
and $\left|-1,0\right\rangle $ states for $\uptau_{L}$=5ns. (b)
Experimentally measured values of $P_{-1,0}$ (circles) as a function
of $\uptau_{L}$ with the theoretical curve given by Eq. (\ref{eq:Population_Eqn})
with $k_{S}=1/0.217\,\mathrm{\mu s}$.\label{fig:Rabi-oscillation-data}}
\end{figure}

To determine the ground state populations, instead of population differences,
we performed an additional experiment; we measured nuclear spin Rabi
oscillations. Fig. \ref{fig:pulse_readout} (rhs) shows the pulse
sequence used for the Rabi measurements. It consists of two consecutive
pulses; one RF Rabi pulse with variable duration $t_{RF}$ and one
MW $\pi$ pulse. The RF pulse exchanges the populations of the $\left|-1,0\right\rangle $
and the $\left|-1,-1\right\rangle $ states and the MW pulse those
of the $\left|-1,0\right\rangle $ and $\left|0,0\right\rangle $
states. Taking into account that $P_{-1,-1}$ vanishes before this
measurement, the readout pulse detects the total population of the
electronic $\left|0\right\rangle $ state as 

\[
P_{m_{S}=0}(t_{RF})=P_{0,-1}+P_{0,+1}+P_{-1,0}\frac{1+\cos\omega_{1}t_{RF}}{2},
\]
where $\omega_{1}$ is the Rabi frequency of the RF field and the
populations $P_{i,j}$ refer to the state before the pulse sequence.
The oscillation amplitude of the Rabi oscillation is thus proportional
to the population $P_{-1,0}$. Fig. \ref{fig:Rabi-oscillation-data}
(a) shows a typical Rabi oscillation for a short laser pulse ($\tau_{L}=5\,\mathrm{ns}$)
and Fig. \ref{fig:Rabi-oscillation-data} (b) shows the observed values
of $P_{-1,0}$ as a function of $\uptau_{L}$. 

\begin{figure}[h]
\centering{}\includegraphics[scale=0.3]{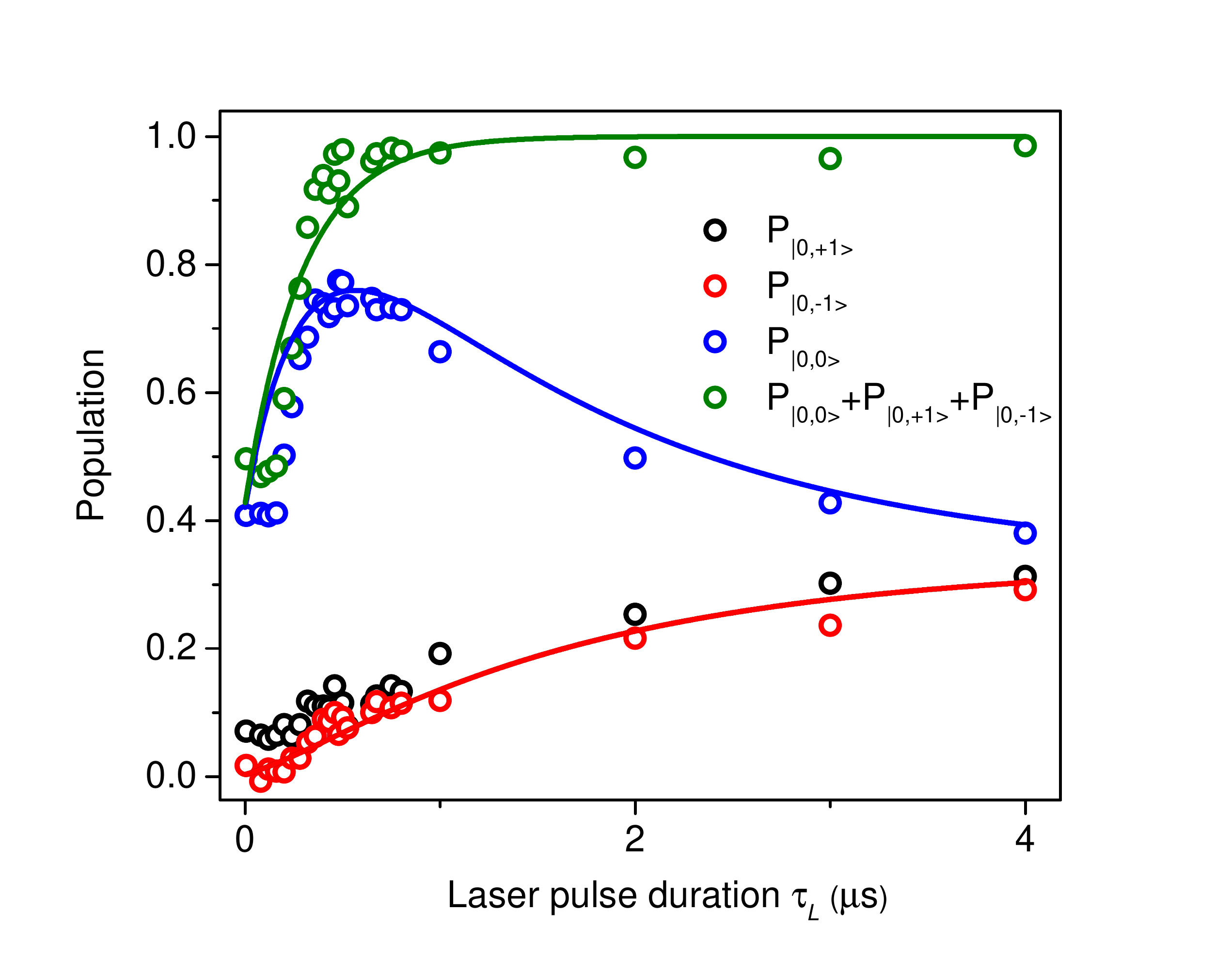}\caption{Measured populations of the three ground states $\left|0,-1\right\rangle $,
$\left|0,+1\right\rangle $, $\left|0,0\right\rangle $ and their
sum as a function of the laser pulse duration $\uptau_{L}$, together
with the fitted curves. \label{fig:Populations}}
\end{figure}

Once $P_{-1,0}$ is determined, it is straightforward to calculate
$P_{0,0}$ using the simple relationship $A_{0}(\uptau_{L})+P_{-1,0}(\uptau_{L})=P_{0,0}(\uptau_{L})$.
Fig. \ref{fig:Populations} shows all three populations $P_{0,0}$,
$P_{0,-1}$ and $P_{0,+1}$ for the initial condition $\vec{P}=\frac{1}{3}(0,0,1,0,0,1,0,0,1)$
as a function of the laser pulse duration $\uptau_{L}$. The experimental
data were normalized by considering that the sum of all nine populations
must add up to 1. The population of the $\left|0,0\right\rangle $
state increases rapidly, with a rate $k_{S}$, until it goes through
a maximum. On a longer time scale determined by $k_{I}$, all three
populations tend towards the equilibrium value of 1/3. The experimental
data for the individual nuclear spin sub-states, as well as the total
population of the $m_{S}=0$ state were fitted with the theoretical
expressions of Eq. (\ref{eq:Population_Eqn}). The resulting values
of the time constants were $1/k_{S}=0.29\pm0.02\mu s$ and $1/k_{I}=4.7\pm0.4\mu s$.
The fact that the effect of the laser pulse illumination started with
a constant time delay in the experimental data, has been taken care
of by adding a constant value $\uptau_{d}$ with $\uptau_{L}$ in
the rate equation model. We estimated $\uptau_{d}=45$ns. The maximum
population (77.8 \%) of the $\left|0,0\right\rangle $ state is obtained
for a pulse duration $\uptau_{L}\approx0.48$ $\mathrm{\mu s}$. For
the $m_{I}=\pm1$ states, the amplitudes $A_{-1}$ and $A_{+1}$,
represent directly the populations $P_{0,-1}$ and $P_{0,+1}$, as
the population of the $\left|-1,-1\right\rangle $ and $\left|-1,+1\right\rangle $
states vanish. 

Different mechanisms lead to loss of nuclear spin polarization during
laser excitation \cite{NeumannScience542,PhysRevLett.111.067601}.
For instance, under laser illumination, hyperfine-induced electron-nuclear
spin flip-flop processes have a dominant effect on the nuclear spin
lifetime $T_{1}$ when the applied magnetic field is such that the
system is close to the excited-state level anticrossing point (LAC).
In addition, coupling of the electronic and nuclear spin with phonons
can lead to depolarization of the nuclear spin. In our case, since
we performed experiments at the field of 6.1 mT which is well below
the LAC ($\approx\pm50$ mT) and spin-orbit coupling has negligible
effect in the $m_{S}=0$ state, the above mentioned reasons have negligible
influence on $T_{1}$. In case of spin flip-flop process, $k_{I}$
includes only the transitions with selection rule $\Delta m_{I}=\pm1$.
Comparing the experimental data with both models, we found that the
data fits better to the model where flips with $\Delta m_{I}=\pm1$
and $\pm2$ occur. We assume that a dominant contribution to nuclear
depolarization is the different energy eigenstates in the ground and
excited state: The optical excitation thus projects the nuclear spin
eigenstates of the electronic ground state onto a superposition of
nuclear spin states in the electronically excited state. Another contribution
is the conversion of the NV$^{-}$ centers to NV$^{0}$ with increasing
laser pulse duration \cite{Aslam_NJP}. The two charge states of the
NV center have different hyperfine coupling and different nuclear
quadrupole interactions. This changes the energy splitting between
the nuclear sublevels and the nuclear spin eigenstates. Accordingly,
a change of charge state may be accompanied by a loss of spin polarization.
In addition, the measured value of $T_{1}$ in the NV$^{0}$ state
is significantly shorter than in the NV$^{-}$ state \cite{Waldherr_PhysRevLett.106.157601,chen2015spin,loretz2016optical}.
The parameter $k_{I}$ summarizes this overall depolarization rate
of the nuclear spin for the given laser irradiation.

\begin{figure}[h]
\centering{}\includegraphics[width=1\columnwidth]{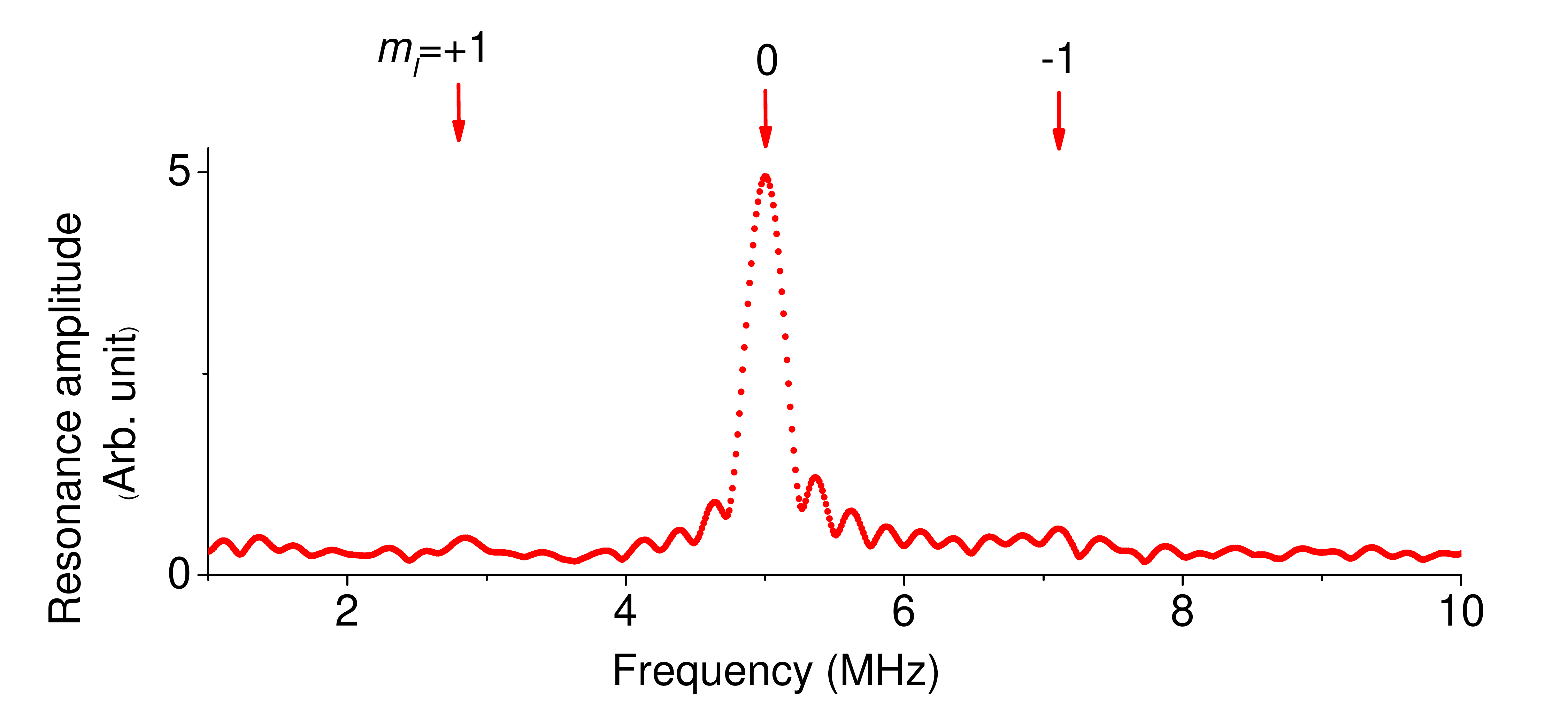}\caption{Experimental spectrum showing the amplitudes $A_{0}$ and $A_{\pm1}$
measured using the read-out procedure shown in Fig. \ref{fig:pulse_readout},
after applying the microwave pulses MW3 and MW4 shown in Fig. \ref{fig:pulses}.
\label{fig:spectrum2}}
\end{figure}

To further improve the purity of the $\left|0,0\right\rangle $ state
in the $m_{S}=0$ subspace, we used two selective microwave $\pi$
pulses MW3 and MW4, both with the same pulse duration of 3.874 $\mu$s,
to transfer the population from the $\left|0,-1\right\rangle $ to
the $\left|+1,-1\right\rangle $ state and from the $\left|0,+1\right\rangle $
to the $\left|+1,+1\right\rangle $ state, respectively. The resulting
spin configuration was read out by performing free-induction-decay
measurements of the electronic spin. The pulse sequence containing
the MW3 and MW4 pulses is shown in the $2^{nd}$ part of Fig. \ref{fig:pulses}
and the FID read out pulses are shown in Fig. \ref{fig:pulse_readout}.
Fig. \ref{fig:spectrum2} shows the resulting spectrum, which contains
a single distinct peak indicating the population of the $\left|0,0\right\rangle $
state. The measured amplitudes show that $>96\%$ of the population
of the $m_{S}=0$ subspace is in the $\left|0,0\right\rangle $ state. 

Apart from the optically induced nuclear spin relaxation, the attainable
polarization is also limited by experimental imperfections. In order
to study the effect of imperfection of the pulse length on the attainable
purity of the $\left|0,0\right\rangle $ state, we have calculated
the derivative of $P_{0,0}$ with respect to each pulse angle while
keeping the other three angles fixed at $\pi$. We obtained $\Delta P_{0,0}=\frac{(\Delta\theta_{i})^{2}}{4}$.
The uncertainty in the pulse angles of the MW and RF pulses is 1.5\%
and 3.2\% which leads to respectively 0.56\% and 2.56\% loss in the
final purity of the $\left|0,0\right\rangle $ state. In addition,
an imperfect initial state influences the achievable purity. If the
first laser pulse initializes the system to a state $\vec{P}=[P_{1},P_{2},1-(P_{1}+P_{2}),0,0,0,0,0,0)$,
the final polarization $P_{0,0}$ becomes $P_{0,0}=0.7858-0.029(P_{1}+P_{2})$
with the boundary conditions $0\leq P_{1}\leq1$, $0\leq P_{2}\leq1$
and $P_{1}+P_{2}\leq1$. The relation indicates that the purity of
$\left|0,0\right\rangle $ state is maximum when $P_{1}=P_{2}=0$
whereas the purity decreases as $P_{1}$ and $P_{2}$ increase. The
average absolute deviation from the mean value of the experimentally
measured quantities $P_{1}$ and $P_{2}$ are 1.65\% and 1.32\% respectively,
which corresponds to 0.086\% loss in the achievable purity of the
$\left|0,0\right\rangle $ state.

\section{Conclusion}

To summarize, we have employed a procedure for the purification of
the quantum state of a two-qubit system associated with a single NV
center in diamond. Laser pulses can initialize the electronic spin.
To initialize also the nuclear spin, we first initialized the electron
spin, swapped the states of the electronic and the nuclear spin and
then again initialized the electronic spin, which is quite similar
to the sequence used by Pagliero et al. \cite{APL(105)242402}. However,
the present work provides a detailed understanding of the population
transfer between the relevant states under optical illumination and
a qualitative estimation of the occupation probabilities of different
individual quantum states. The corresponding values were obtained
by performing partial quantum state tomography. We were able to establish
an effective rate equation model which is in excellent agreement with
the experimental results. The interpretation of the results using
the formulated theory enabled us to determine the relevant rate constants
and the optimal duration of the laser pulse. Thus, the optimal sequence
allows one to prepare an arbitrary state of the system with high purity,
which is essential for coherently controlled experiments on NV centers
in diamond. Although the resultant purity is limited by the initial
laser-induced electronic polarization of $m_{S}=0$ state and nuclear
depolarization, the procedure offers the benefit that it can be implemented
with flexible magnetic field values.

Finally, we have improved the purity of the $\left|0,0\right\rangle $
state in the $m_{S}=0$ subspace. Enhancing the occupation probability
of the target input state in a certain computational space can result
in a better performance in experiments like quantum gate operations.
For instance, in an experiment where dynamical decoupling was applied
to implement protected operation of a controlled rotation gate in
a definite subspace \cite{Jingfu_2015}, an enhanced purity of the
input state increases the signal-to-noise ratio.
\begin{acknowledgments}
We gratefully acknowledge useful discussions with K. Rama Koteswara Rao, Manpreet Kaler and Fabian Lehmann. We also sincerely thank Nabeel Aslam and Matthias Pfender for useful discussions. This work was supported by the DFG through grants 192/19-2, 192/34-1 and by the MERCUR foundation through grant Pr-2013-0003. We acknowledge financial support by Deutsche Forschungsgemeinschaft and TU Dortmund Technical University within the funding programme Open Access Publishing.
\end{acknowledgments}

\bibliographystyle{apsrev}
\bibliography{PSP_bib}

\section{Appendix}

The matrix $M(k_{S},k_{I})$ describing the effect of the laser pumping
on the electronic and nuclear spin is, $M(k_{S},k_{I})=\left[\begin{array}{ccccccccc}
-2k_{I} & k_{I} & k_{I} & k_{S} & 0 & 0 & k_{S} & 0 & 0\\
k_{I} & -2k_{I} & k_{I} & 0 & k_{S} & 0 & 0 & k_{S} & 0\\
k_{I} & k_{I} & -2k_{I} & 0 & 0 & k_{S} & 0 & 0 & k_{S}\\
0 & 0 & 0 & -k_{S} & 0 & 0 & 0 & 0 & 0\\
0 & 0 & 0 & 0 & -k_{S} & 0 & 0 & 0 & 0\\
0 & 0 & 0 & 0 & 0 & -k_{S} & 0 & 0 & 0\\
0 & 0 & 0 & 0 & 0 & 0 & -k_{S} & 0 & 0\\
0 & 0 & 0 & 0 & 0 & 0 & 0 & -k_{S} & 0\\
0 & 0 & 0 & 0 & 0 & 0 & 0 & 0 & -k_{S}
\end{array}\right]$ where the population vector is depicted with the order ($m_{S},m_{I}$)
= (0,-1; 0,+1; 0,0; -1,-1; -1,+1; -1, 0; +1,-1; +1,+1; +1,0).

For an initial state $\vec{P}=(P_{1},P_{2},P_{3},P_{4},P_{5},P_{6},P_{7},P_{8},P_{9})$
and after the second laser pulse of duration $\uptau_{L}$, the resulting
population vector is 

\begin{align*}
\vec{P} & =\frac{1}{3}(1+\frac{ae^{-3k_{I}\uptau_{L}}+3be^{-3k_{S}\uptau_{L}}}{(3k_{I}-k_{S})},1+\frac{3ce^{-k_{S}\uptau_{L}}+fe^{-3k_{I}\uptau_{L}}}{(3k_{I}-k_{S})},\\
 & 1+\frac{3ge^{-k_{S}\uptau_{L}}+he^{-3k_{I}\uptau_{L}}}{(3k_{I}-k_{S})},3P_{4}e^{-k_{S}\uptau_{L}},3P_{5}e^{-k_{S}\uptau_{L}},\\
 & 3P_{6}e^{-k_{S}\uptau_{L}},\\
 & 3P_{7}e^{-k_{S}\uptau_{L}},3P_{8}e^{-k_{S}\uptau_{L}},3P_{9}e^{-k_{S}\uptau_{L}}),\\
\end{align*}

where 

$a=k_{S}-3k_{I}P_{3}+6k_{I}P_{1}-3k_{I}P_{2}-3k_{S}P_{1}-3k_{S}P_{4}-3k_{S}P_{7}$,

$b=k_{I}P_{3}-k_{I}+k_{I}P_{1}+k_{I}P_{2}+k_{S}P_{4}+k_{S}P_{7}$,

$c=k_{I}P_{3}-k_{I}+k_{I}P_{1}+k_{I}P_{2}+k_{S}P_{5}+k_{S}P_{8}$,

$f=k_{S}-3k_{I}P_{3}-3k_{I}P_{1}+6k_{I}P_{2}-3k_{S}P_{5}-3k_{S}P_{2}-3k_{S}P_{8}$,

$g=k_{S}-k_{I}-k_{S}(P_{5}+P_{4}+P_{7}+P_{8})+(k_{I}-k_{S})(P_{3}+P_{1}+P_{2})$
and

$h=6k_{I}P_{3}-2k_{S}+3k_{S}(P_{5}+P_{4}+P_{7}+P_{8})-3(k_{I}-k_{S})(P_{1}+P_{2})$.
\end{document}